# Harnessing subcellular-resolved organ distribution of cationic copolymer-functionalized fluorescent nanodiamonds for optimal delivery of active siRNA to a xenografted tumor in mice

Sandra Claveau[§a,b], Marek Kindermann[§c,d], Alexandre Papine[§e], Zamira V. Díaz-Riascos[f,g], Xavier Délen[h], Patrick Georges[h], Roser López-Alemany[i], Òscar Martínez Tirado[i], Jean-Rémi Bertrand[b], Ibane Abasolo[f,g], Petr Cigler[c], and François Treussart*[a]

[a.] Université Paris-Saclay, ENS Paris-Saclay, CNRS, CentraleSupélec, LuMIn, 91190 Gif-sur-Yvette, France. E-mail : francois.treussart@ens-paris-saclay.fr
[b.] Université Paris-Saclay, Institut Gustave Roussy, CNRS, Metabolic and Systemic Aspects of Oncogenesis (METSY), 94805 Villejuif, France.
[c.] Institute of Organic Chemistry and Biochemistry of the Czech Academy of Sciences, 166 10 Prague, Czech Republic.
[d.] Department of Chemical Engineering, University of Chemistry and Technology, 166 28 Prague, Czech Republic.
[e.] IMSTAR S.A., 75014 Paris, France.
[f.] Drug Delivery & Targeting, Functional Validation & Preclinical Research (FVPR), CIBBIM-Nanomedicine, Vall d'Hebron Institut de Recerca (VHIR), Universitat Autònoma de Barcelona (UAB), 08035 Barcelona, Spain.
[g.] Networking Research Center on Bioengineering, Biomaterials and Nanomedicine (CIBER-BBN), 08035 Barcelona, Spain.
[h.] Université Paris-Saclay, Institut d'Optique Graduate School, CNRS, Laboratoire Charles Fabry, 91127 Palaiseau, France
[i.] Sarcoma Research Group, Oncobell Program, Bellvitge Biomedical Research Institute (IDIBELL), 08908 L'Hospitalet de Llobregat, Barcelona, Spain

§ These authors contributed equally to this work.

## Abstract

Diamond nanoparticles (nanodiamonds) can transport active drugs in cultured cells as well as *in vivo*. However, in the latter case, methods allowing to determine their bioavailability accurately are still lacking. Nanodiamond can be made fluorescent with a perfectly stable emission and a lifetime ten times longer than the one of tissue autofluorescence. Taking advantage of these properties, we present an automated quantification method of fluorescent nanodiamonds (FND) in histological sections of mouse organs and tumor, after systemic injection. We use a home-made time-delayed fluorescence microscope comprising a custom pulsed laser source synchronized on the master clock of a gated intensified array detector. This setup allows to obtain ultra-high-resolution images 120 Mpixels of whole mouse organs sections, with subcellular resolution and single-particle sensitivity. As a proof-of-principle experiment, we quantified the biodistribution and aggregation state of new cationic FNDs able to transport small interfering RNA inhibiting the oncogene responsible for Ewing sarcoma. Image analysis showed a low yield of nanodiamonds in the tumor after intravenous injection. Thus, for the *in vivo* efficacy assay we injected the nanomedicine into the tumor. We achieved a 28-fold inhibition of the oncogene. This method can readily be applied to other nanoemitters with ≈100 ns lifetime.

## 1. Introduction

The *in vivo* efficacy of drugs is largely dependent on their bio-availability and -distribution after body administration. One strategy to optimize the tissue distribution relies on delivering the drugs with nanoparticles. It led to the emergence of the field of nanomedicine about 40 years ago. The quantitative assessment of nanomedicine distribution *in vivo* is most often done with apparatus dedicated to whole animal body imaging of radioelement, bioluminescence or fluorescence reporters (especially in the near infrared wavelength tissue transparency region)[1].

To date, the spatial resolution achieved by these apparatus does not allow to resolve single cell, reaching at best 20 µm (with the fluorescence molecular tomograph IVIS Spectrum, PerkinElmer, USA). Such low resolution makes the identification of the delivery and elimination pathways difficult, and motivates the development of cellular and sub-cellular resolved methods. The latter requires larger magnification and can be carried out either *in vivo* by intravital microscopy[2] and endoscopy[3], or *ex vivo* on tissue sections. In order to achieve a reliable quantification of the nanomedicine organ distribution at subcellular resolution, a large number of fields of views (FOV, typically of 100 µm in size) need to be recorded and analyzed, to cover a whole section, requiring an automatization of the process. Moreover, for its efficient detection the nanomedicine should possess a sufficiently large contrast and specificity in the imaging modality used.

While advanced technologies as time-of-flight secondary ion mass spectroscopy imaging offer unique specificity along with sub-micrometer resolution[4], they are usually not carried out on whole organ section, as it can be more easily done by optical microscopy. In this domain, hyperspectral dark-field microscopy[5] has shown its ability to detect strongly scattering nanoparticles (*e.g.* gold nanoparticles) in *ex vivo* tissue section at the single particle level with subcellular resolution, allowing precise quantification of their biodistribution[6]. While it is a powerful method when dealing with strongly scattering material, it may be less efficient with other nanovector materials.

Among them, diamond nanocrystal (nanodiamond, size 5-50 nm) has been shown to be an efficient delivery agent of macromolecules such as anticancer compounds or siRNA into cells in culture or small organisms, and that can be traced by various remarkable modalities on unlimited timescale thanks to its perfectly stable structure[7–9]. The first modality is associated to the Raman scattering signature of diamond that was used in particular to evaluate the overall tissue distribution after whole organ digestion[10]. The second modality, which has popularized nanodiamonds these past fifteen years, is the possibility to entitle them with a perfectly photostable fluorescence by generating nitrogen-vacancy (NV) defects center within the diamond lattice, creating fluorescent nanodiamonds (FND)[8]. Upon green laser excitation, NV center emits in the 600-750 nm wavelength range and possess at least two additional unique properties of interest for tissue distribution. First of all, the negatively charged $NV^-$ has an optically detectable electron spin resonance that is largely harnessed for sensitive magnetometry and quantum sensing[11], and was also used to implement a magnetic resonance imaging scheme and identify FND in tissue[12], but at a spatial resolution limited to 100 µm. Secondly, the insertion of the NV center in the subwavelength scale sized nanodiamond lattice, results in an increase in its emission lifetime (≈20-40 ns) compared to bulk diamond environment (≈11 ns), due to a smaller local density of electromagnetic states[13]. FND emission lifetime is therefore about one order of magnitude longer than the one of tissue autofluorescence (≈3 ns[14]), allowing an efficient filtering of FND by time-gated detection in a sub-nanosecond pulsed excitation scheme[15]. This strategy was already exploited to monitor tissue regeneration by tracking FND-labelled stem cells in mouse lung section with single FND resolution[16], but it was not extended yet to their quantitative biodistribution.

Here we report the development of an automated imaging and analysis biodistribution pipeline able to (i) quantify the FND content in histological sections of mouse main organs after systemic injection, and to (ii) infer the cellular type thanks to histological staining, subcellular resolution and single particle resolution. We apply this method to the determination of the optimal delivery route by nanodiamonds of an anticancer compound to a tumor xenografted on mice. In the continuity of our previous work on the treatment of Ewing sarcoma (ES)[17,18] –a cancer in which cell proliferation is driven by the expression of a junction oncogene – we use a small interfering RNA (siRNA) to inhibit the oncogene. We also improve the siRNA electrostatic binding to FND using a linear cationic copolymer covalently grafted from the FND surface ($Cop^+$-FND). This conjugate enables to protect siRNA against degradation in physiological environment and to deliver it efficiently to the cells. We validate the *in vitro* efficacy of the new $Cop^+$-FND:siRNA to inhibit the oncogene in ES cell in culture, with an efficacy larger than the one reached previously with FND complexed with poly(ethyleneimine) (FND@PEI)[17,19]. Using our automatized biodistribution pipeline, we map the $Cop^+$-FND:siRNA bioavailability and quantify its tissue distribution *in vivo* after systemic intravenous (i.v.) injection. Following the results of this injection, we design a new treatment strategy, relying on intratumoral (i.t.) administration, leading to efficient inhibition of the oncogene and a tumor growth rate reduction. Our results demonstrate that the automated quantitative biodistribution method we developed for FND injected in mice, is a valuable tool to design a treatment strategy. We show that it can also provide insights on nanodiamond cellular fate and elimination pathways thanks to its sub-cellular resolution and high sensitivity of nanoparticle detection.

## 2. Results and discussion

### 2.1. $Cop^+$-FND:siRNA optimal design and its oncogene inhibition efficacy in human Ewing Sarcoma cultured cells

In early studies, we used FND@PEI to deliver siRNA to cultured ES cells and we obtained a significant reduction of the expression of the oncogene responsible for the cell proliferation[17,18]. Here we replaced this strategy with $Cop^+$-FND, a new generation of a cationic copolymer interface covalently grafted from the FND surface. We first investigated the size and charge of this novel Cop$^+$-FND:siRNA complex. Our copolymer consists of two components (Scheme 1). As the cationic comonomer we chose (2-dimethylaminoethyl) methacrylate (DMAEMA), which homopolymer (pDMAEMA) and various copolymers have been described as effective vectors for siRNA delivery[20]. Considering a previous study focusing on transfection with the charge diluted copolymers of DMAEMA with electroneutral comonomers ethoxytriethylene glycol methacrylate and *N*-vinyl-pyrrolidone[21], we implemented 21 mol. % of *N*-(2-hydroxypropyl) methacrylamide (HPMA) monomer in the copolymer structure (the exact composition of the copolymer used in this study, together with detailed chemical and structural characterization of the particles are detailed in Ref.[22]). To obtain robust data, we independently synthesized three different batches of Cop$^+$-FND and analyzed their colloidal properties. The samples showed a great batch-to-batch reproducibility with narrow distribution of *Z*-average diameter 128.2 ± 8.7 nm (measured by dynamic light scattering, DLS), an apparent ζ-potential of 46.0 ± 3.7 mV (measured by electrophoretic light scattering, ELS), and sample conductivity 4.2 ± 1.8 µS/cm in nuclease free water (pH 5.1). All these values are mean ± standard deviation, over the three independent batches, and the measurement procedure for one batch is described in the Experimental section. The corresponding transmission electron microscopy (TEM) micrographs (Supporting Fig. S1a,b) show the characteristic sharp shape of FND particles, which is not changed after the modification by copolymer. The Cop$^+$-FND particles are not aggregated onto the TEM grid, which is consistent with the DLS measurement.

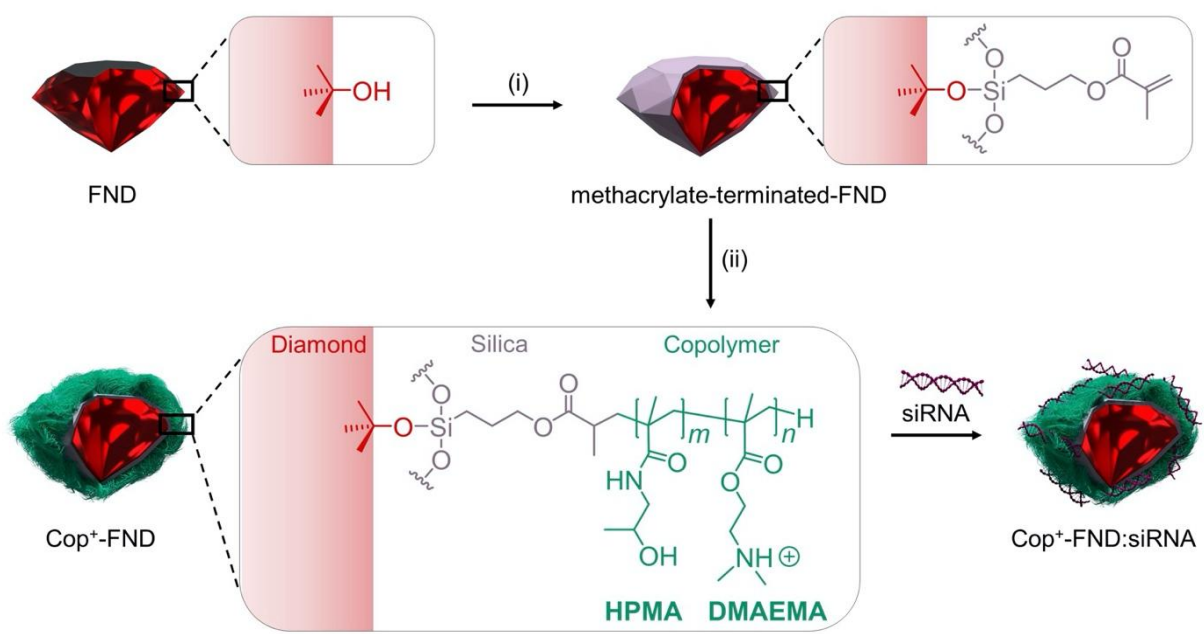

**Scheme 1. Different steps of synthesis of the cationic copolymer functionalized fluorescent nanodiamond Cop$^+$-FND**: (i) growth of methacrylate groups from FND surface; (ii) addition of HPMA and cationic DMAEMA copolymers. Bottom right: schematics of Cop$^+$-FND:siRNA conjugate.

Here we replaced this strategy with Cop$^+$-FND, a new generation of a cationic copolymer interface covalently grafted from the FND surface. We first investigated the size and charge of this novel Cop$^+$-FND:siRNA complex. Our copolymer consists of two components (Scheme 1). As the cationic comonomer we chose (2-dimethylaminoethyl) methacrylate (DMAEMA), which homopolymer (pDMAEMA) and various copolymers have been described as effective vectors for siRNA delivery[20]. Considering a previous study focusing on transfection with the charge diluted copolymers of DMAEMA with electroneutral comonomers ethoxytriethylene glycol methacrylate and *N*-vinyl-pyrrolidone[21], we implemented 21 mol. % of *N*-(2-hydroxypropyl) methacrylamide (HPMA) monomer in the copolymer structure (the exact composition of the copolymer used in this study, together with detailed chemical and structural characterization of the particles are detailed in Ref.[22]). To obtain robust data, we independently synthesized three different batches of Cop$^+$-FND and analyzed their colloidal properties. The samples showed a great batch-to-batch reproducibility with narrow distribution of *Z*-average diameter 128.2 ± 8.7 nm (measured by dynamic light scattering, DLS), an apparent ζ-potential of 46.0 ± 3.7 mV (measured by electrophoretic light scattering, ELS), and sample conductivity 4.2 ± 1.8 µS/cm in nuclease free water (pH 5.1). All these values are mean ± standard deviation, over the three independent batches, and the measurement procedure for one batch is described in the Experimental section. The corresponding transmission electron microscopy (TEM) micrographs (Supporting Fig. S1a,b) show the characteristic sharp shape of FND particles, which is not changed after the modification by copolymer. The Cop$^+$-FND particles are not aggregated onto the TEM grid, which is consistent with the DLS measurement.

For the siRNA, we considered a siRNA sequence directed against *EWS-FLI1* junction oncogene (siAS), as in our previous work[17,18,23]. This oncogene is involved in the vast majority of Ewing sarcoma, a young adult bone and soft tissue cancer[24] with bad prognosis. Using siAS we determined the optimal Cop$^+$-FND:siAS mass ratio to be 25:1 by varying the amount of Cop$^+$-FND at a fixed concentration of siAS in water at 25°C. The mass ratio 25:1, in combination with the optimized protocol, provided very stable colloidal suspensions and strongly limited aggregation at all stages of the preparation. Formation of Cop$^+$-FND:siAS complex from Cop$^+$-FND in water led to a slight increase of the *Z*-average diameter to 150.9 ± 6.2 nm and a decrease of the apparent ζ-potential down to 38.4 ± 0.8 mV, due to charge compensation (Supporting Table S1 & Fig. S1c,d).

Then, after Cop$^+$-FND:siAS formation in water at 25°C, we tested its colloidal stability in serum free DMEM, 10% and 100% FCS at 37°C (Fig. 1a). While the diluted FCS reveals a stabilization effect thanks to the presence of serum proteins, serum free DMEM strongly destabilizes the particles leading to rapid aggregation[25]. Adsorbing serum proteins in 100% FCS caused an almost two-fold increase of the diameter and a slow aggregation (Supporting Table S1 & Fig. S1c,d).

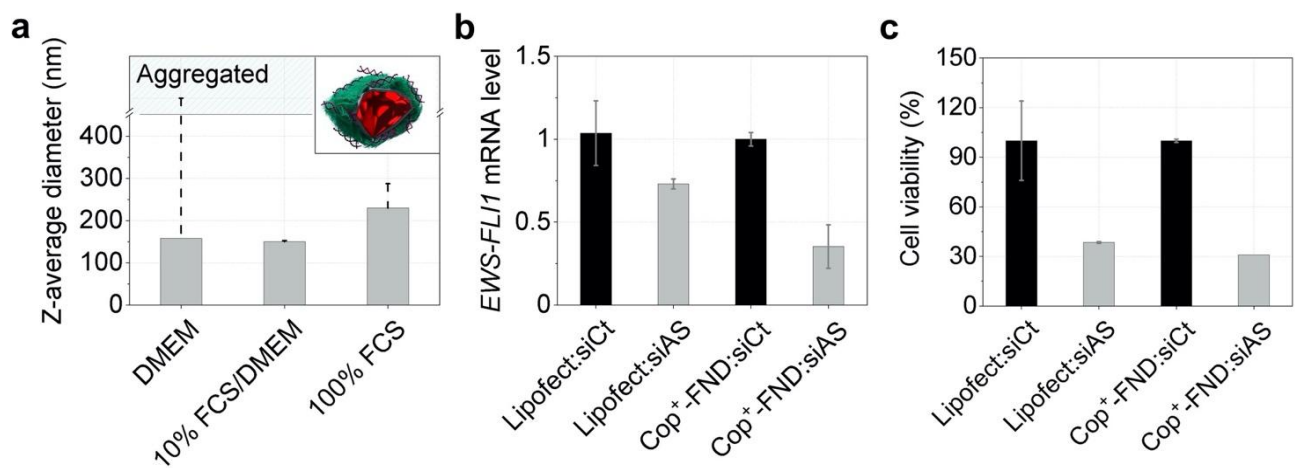

**Figure 1. Cop$^+$-FND:siRNA conjugate characterization: size and *in vitro* oncogene inhibition efficacy.** A) Hydrodynamic size (*Z*-average) measurement by DLS of the Cop$^+$-FND:siAS complex at a diamond:siRNA mass ratio of 25:1, in different medium conditions at 37 °C (Inset: scheme of the complex). All measurements were done with Cop$^+$-FND batch #1 and repeated ten times (during approximately 20 min). Each bar represents the first measurement and the top part of the superimposed dashed line the 10$^{th}$ one. More detailed data (raw DLS intensity autocorrelation functions and intensity size distributions) are shown in Table S1 and Figure S1c,d, in the ESI. b) *In vitro* inhibition of *ESW-FLI1* mRNA expression in A673 ES cells in culture, using 100 nM siRNA, which corresponds to 1.34 µg siRNA/mL and 33.4 µg Cop$^+$-FND/mL at 25:1 mass ratio (*n*=3 experiments). Measurement by RT-qPCR, 48 h post-treatment. c) Inhibition of the proliferation of A673 cells measured 72 h post-treatment (*n*=2 experiments). For b) and c) data displayed are mean ± s.e.m, except for Cop$^+$-FND:siAS in c) for which a single valid measurement could be done.

aggregation[25]. Adsorbing serum proteins in 100% FCS caused an almost two-fold increase of the diameter and a slow aggregation (Supporting Table S1 & Fig. S1c,d).

We then measured the inhibition efficacy of the siAS vectorized by Cop$^+$-FND in A673 human Ewing sarcoma cells in culture. After a 48 h incubation with the Cop$^+$-FND:siAS, *EWS-FLI1* messenger RNA (mRNA) was extracted and its fraction was measured compared to a reference by RT-qPCR. Fig. 1b shows that *EWS-FLI1* expression in A673 cells treated with Cop$^+$-FND:siAS (25:1 mass ratio) was specifically inhibited compared to a treatment with an irrelevant siRNA (siCt), and to the siAS delivered by Lipofectamine 2000. Inhibition of *EWS-FLI1* was followed by a proliferation capacity of A673 cells decreased to 31%, compared to the cell viability observed in cells treated with Cop$^+$-FND:siCt (Fig. 1c). Moreover, *EWS-FLI1* inhibition is dependent on siAS concentration. In an additional separate experiment we demonstrated 25% inhibition at 30 nM siRNA and 90% for 150 nM siRNA when Cop$^+$-FND:siAS (65:1 mass ratio) was used in the presence of serum in the transfection media (Supporting Fig. S2). The capacity of Cop$^+$-FND to deliver an efficient siRNA to cells in a serum containing medium is a superior advantage for the animal studies we then carried out.

Interestingly, our experiments also showed that the optimal Cop$^+$-FND:siAS mass ratio providing stable colloidal sample depends on the content of salt impurities which differs between siRNA suppliers. We repeatedly obtained the optimal mass ratio 25:1 with siRNA purchased from Sigma Aldrich. Improperly purified siRNA lead to an increase of the Cop$^+$-FND:siRNA mass ratio to 65:1 to maintain colloidal stability; biological activity was unaffected (ESI† Fig. S2). Compared to the optimal estimated ratio of 140:1 in the case of FND@PEI[17], these results indicate that Cop$^+$-FND has a much larger binding capacity to carry siRNA.

## 2.2. High-resolution automatic quantification of Cop$^+$-FND:siRNA in mouse organ sections after intravenous injection

To evaluate the siRNA inhibition efficacy *in vivo* we xenografted an Ewing sarcoma cells tumor in mice, and we first considered i.v. administration of the Cop$^+$-FND:siAS (injection of a siRNA mass of 1 mg/kg). The siRNA delivery efficacy in animals is largely dependent on nanoparticles distribution after administration in blood circulation. Because no specific targeting is associated to Cop$^+$-FND, they are expected to be captured by all cells having endocytosis capacities. However, the *Z*-average size of Cop$^+$-FND:siAS, smaller than 300 nm, lies in the range where the enhanced permeability and retention (EPR) effect[26] may allow them to extravasate from the tumor blood vessels into the tumor microenvironment, leading to their larger accumulation in the tumor. We decided to measure the main organ and tumor FND distribution at high sensitivity and spatial resolution. To this aim we developed an automated quantitative biodistribution measurement pipeline relying on time-gated widefield detection of FND in histological tissue sections.

**2.2.1. Large-scale high-resolution image acquisition with a home-made time-gated microscope.** The heart, the liver, the kidneys, the lungs, the spleen and the tumor were collected in two mice sacrificed 24 hours after i.v. injection, sectioned (4 μm thick), and inserted between a coverslip and a microscope slide for imaging with a home-made time-gated wide field fluorescence microscope, described in details the Experimental section. Briefly, this setup relies on an inverted fluorescence microscope automatic slide scanner, in which we inject the beam of a home-made pulsed laser (wavelength 532 nm) shaped for widefield illumination synchronized with the delayed-detection via an intensified CCD (ICCD) array detector (Fig. 2a). We delayed the detection by 15 ns from the laser excitation pulse, to strongly reduce the autofluorescence compared to the FND emission, which lasts one order of magnitude longer[15] (Fig. 2b). A dedicated software drives the slide scanner displacements and image acquisitions, a single organ section coverage requiring the recording of up to 15,000 individual images (FOV of size ≈110 μm with a ×60 oil-immersion microscope objective). The same software serves to stitch these individual images and form a ≈120 Mpixels high-resolution image (Fig. 2c), containing the millimeter size whole section with a pixel resolution of 200 nm. Note that the intensifier phosphor of the ICCD limits the spatial resolution of the image, spreading photons on ≈3.3 pixels full-width at half-maximum (FWHM) according to the ICCD manufacturer specifications. If we convolute this spreading with (i) the diffraction limit of our optical system (240 nm FWHM, for 700 nm emission wavelength and 1.40 numerical aperture objective) and (ii) the FND size (≈50 nm) we get a point-spread function (spot of a single emitter in the recorded image) of ≈5 pixels FWHM in agreement with our experimental resolution.

**2.2.2. Automatic detection of FND in organ section images.** The detection and quantification of the Cop$^+$-FNDs in the different organ sections were realized thanks to a dedicated image analysis pipeline. Briefly, small elements brighter than the mean image intensity were first identified in FND fluorescence channel, with a Top-hat filter, and a region of interest (ROI) identifier was assigned to each of them. This first image processing did not include any size differentiation, thus a filtering program for every organ has been developed, considering the residual autofluorescence level and the anatomy of each tissue, these parameters being different from one organ to the other (Supporting Table S2). The result of this automatic detection can be seen in Fig. 2d and in Fig. 3a,b for other FOV in liver and tumor. Furthermore, in order to have both histopathological information and Cop$^+$-FND localization, some sections were stained with Hematoxylin/Eosin/Saffron (HES). In that case, a bright-field color image was also acquired in addition to the fluorescence one, using a red-blue-green collimated LED (Fig. 2e). However, despite the time-gating modality that rejects short-lifetime emitted photons, HES staining lead to a too high level of auto-fluorescence, so that stained samples were only used to better understand cellular and sub-cellular localization of Cop$^+$-FNDs as discussed in section 2.2.3.

**2.2.3. FND tissue distribution: quantification and aggregation state.** For the quantification of Cop$^+$-FND organ distribution we only used non-stained sections. To ensure the detection of all FND in the whole depth of the 4 μm thick organ section, we acquire two images (350 ms duration each) at two focusing $z$ depths 3 μm apart. Moreover, even with time-gating detection and absence of staining, the tissue auto-fluorescence could not be totally suppressed to discriminate single FND unambiguously from localized tissue autofluorescence. We therefore also added a bleaching step (same duration of 350 ms) before each $z$-plane acquisition. The bleaching step decreased the auto-fluorescence by about 40%, without affecting Cop$^+$-FND signal, thanks to its perfect photostability. Overall, each FOV is recorded four times. FND identification was then based on a $z$-projection (of the bleaching and of the following steps) and followed by a comparison of the relative intensity of FND candidate spots to the background, before and after the bleaching: only FND spots are expected to keep their intensity constant. Fig. 3a,b left panel shows the raw image before ROI identifications in liver ant tumor sections respectively, in which single and aggregated Cop$^+$-FND appear as white spots, and the corresponding right images show a white line delineating the ROI around identified Cop$^+$-FND.

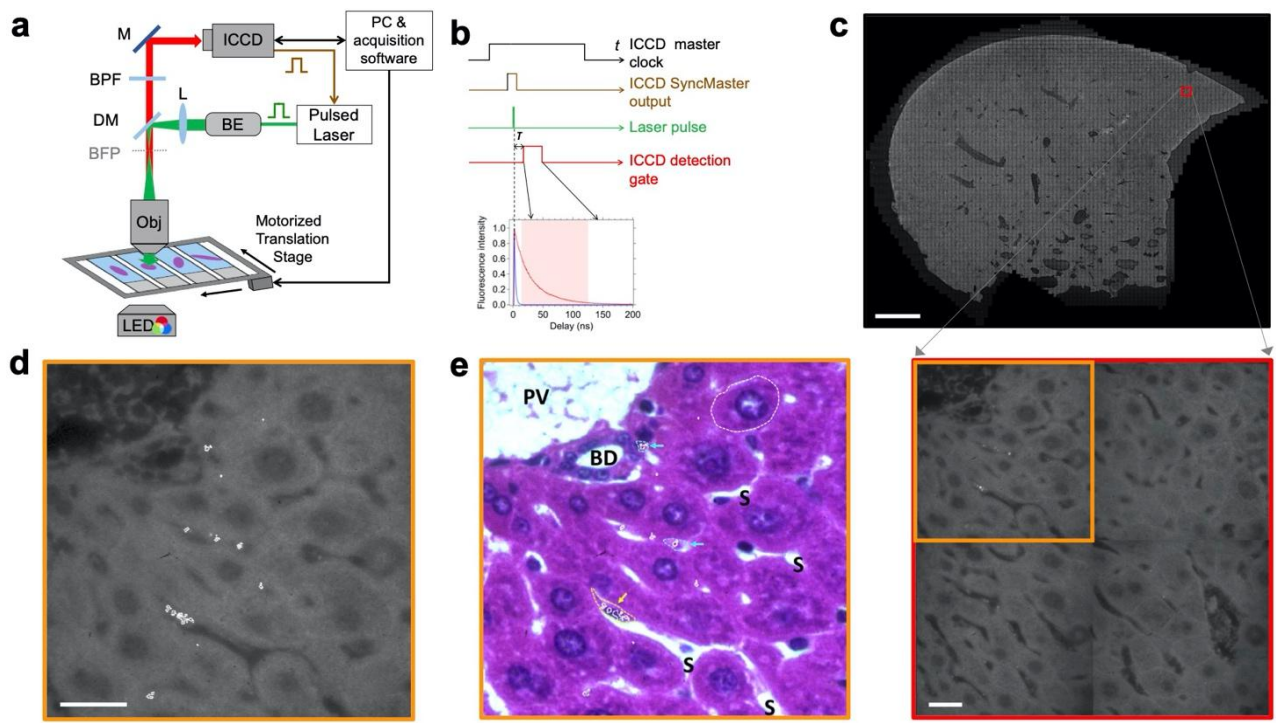

**Figure 2. Automatised time-gated fluorescence and bright-field microscopy setup to screen nanodiamond in mouse organ sections with subcellular resolution.** a) Motorized epifluorescence microscope equipped with a slide holder containing up to 4 slides with organ section preparations. BE: beam expander; DM: dichroic mirror; L: lens; BFP: microscope objective back focal plane; BPF: bandpass filter; ICCD: intensified charge coupled device array detector. internal clock. b) Chronogram of the acquisition sequence. ICCD clock signal (SyncMaster Output) serves as a master to triggers the laser pulse. The fluorescence detection takes place with an adjustable delay relative to the pulse emission, and during a given gate duration (light red shaded area in the fluorescence decay curve). For each field of view, a brightfield image is also acquired on the same array detector, from three consecutive illuminations by a red, blue and green LED integrated in a condenser. A motorized translation stage, controlled by the

acquisition software, ensures the motion from one field of view to the adjacent one, as well as from one slide to the next one. The areas covered by the organ slices in each slide, and a set of focus landmarks, have to be defined prior to the acquisition, which is then fully autonomous. c) Top: mosaic of 6500 field-of-view (FOV) of fluorescence images acquired from a HES-stained liver section. Scale bar: 1 mm. Bottom: zoom on 4 adjacent FOV of a region framed in red. The FOV surrounded by an orange frame displays bright spots corresponding to FND. Scale bar: 20 µm. d) Bright-field image (histopathology view) of the same FOV surrounded by orange, and resulting from the composition of the three LED acquisitions. Cell nuclei appear darker. e) Fluorescence image of the same FOV as in d) on top of which regions of interest (ROI) where FND have been automatically detected are delimited by white lines. Scale bar: 20 µm. These FND-containing ROI are superimposed to the bright field d) image. In d) the arrows point at cells containing FND. The two blue arrows point at what may be Kupffer cells while the yellow arrow point at an endothelial cell. S: sinusoids; BD: bile duct; PV: portal vein.

Our automatic and high-throughput detection allowed to quantify the $Cop^+$-FND in each organ, by extracting different parameters. The first one is the level of fluorescence intensity per organ section area (Fig. 3c), which is related to the number of $Cop^+$-FND; it would be proportional to it if each single FND had the same fluorescence level, which is not the case. The organ section displaying the highest fluorescence intensity per unit area was the spleen (1.71±0.57 count/µm$^2$), closely followed by the liver (1.37±0.64 count/µm$^2$). The lung, which was the organ in which the largest number of ROIs was detected, comes third (0.75±0.35 count/µm$^2$). Finally, the kidneys (0.2±0.04 count/µm$^2$), the tumor (0.08±0.01 count/µm$^2$) and the heart (0.05±0.01 count/µm$^2$) presented much smaller accumulation of $Cop^+$-FND. A second meaningful parameter is the ROI area representing the aggregation state of $Cop^+$-FND shown on Fig. 3d. Three groups can be distinguished. The one of the liver, which separates from the other organs by a median ROI size of 4.2±3.9 µm$^2$ and a very large dispersion with a maximum at almost 11 µm$^2$. The second group is made of the lung, spleen and kidney, having similar large ROI median sizes and dispersions of about 2.7±2 µm$^2$. The third group of organs in ROI area consists in the heart and the tumor, the latter presenting the lowest median value which indicates a minor aggregation state in this tissue. This might be the sign of a lower cellular uptake efficiency by the large macropinocytotic vesicles (size >1 µm) in tumors compared to other organs like the liver. In previous studies in cultured cells we have identified that the macropinocytotic compartments (as opposed to endosomal ones, sizes ≈100-500 nm) allowed the release of siRNA in the cytosol where these molecules are active[18,23].

The liver, kidneys and spleen are organs which function is to filtrate the blood and remove undesirable metabolites, xenobiotics and participate to the destruction of impaired blood cells. This function leads to all body blood filtration and may be responsible for the larger accumulation of nanoparticles. However, considering that $Cop^+$-FND accumulates and aggregates also in the lung in addition to the filtration organs, one cannot exclude that the nanoparticle aggregation takes place in the blood stream, leading to $Cop^+$-FND entrapment in alveolar capillaries. The broad distribution of ROI sizes in the lung is most probably due to the diversity of sizes of capillaries. Regarding the tumor, our observation of a low accumulation is in agreement with a recent meta-analysis concluding that less than 1% in mass of nanoparticles administered systemically end up in the tumor target[27], but the fact that the amount found in the tumor is up to 17 fold smaller compared to the one found in the liver tends to contradict the EPR "passive" targeting effect of nanomedicine[26]. However, it has been pointed out that EPR is heterogeneous and can be affected by physiological and pathological effects[28]. Moreover, our observations have been done at a single time-point of 24 h after intravenous injection. Therefore, our data does not allow to discriminate between either an ineffective EPR effect, or a fast journey of the $Cop^+$-FND through the tumor, thanks to a low aggregation state in this organ. Only additional experiments, at time points shorter than 24 h, would allow to conclude on the dominant phenomenon.

**2.2.4. FND cellular localization in liver histological sections.**

To investigate in which compartments of the liver $Cop^+$-FND end up, we then took advantage of high magnification of individual image constituting the mosaic and of the bright field acquisition of HES stained sections. Fig. 2e shows examples of $Cop^+$-FND ROI superimposed to HES images. FND were found in majority internalized in endothelial cells bordering the sinusoids, and second, most probably inside Kupffer cells. The latter are liver-resident macrophages part of the mononuclear phagocyte system and essential to maintain liver homeostasis and scavenge xenobiotics and cell debris[29]. A staining specific of macrophages like F4/80[30] would allow to unambiguously distinguish Kupffer cells from endothelial cells, so that we could have an accurate quantification of each cell type. The capture by Kupffer cells, if it is confirmed, would not be favourable to the delivery of nanomedicine into the target tumour, but it is not the sole mechanism responsible for the low yield of nanoparticles ending up in tumor after systemic administration, as recently reported in models where Kupffer cells were removed[31]. The interaction with other organs than the liver, like the spleen, need also to be tuned to achieve a larger tumor delivery.

Let us further point out that the phagocytosis of nanodiamonds by Kupffer cells was already reported a decade ago[10], and that these cells are expected to play an important role in the elimination of these particles from the organism. However, we did not detect any nanodiamond in hepatocytes 24 h after administration, which indicates that we either did not wait long enough for their passing from the Kupffer cells to the hepatocyte before reaching the bile duct and being eliminated in the feces, or that they are sequestered for a longer period and might not be eliminated at all. In any case, the important phagocytosis of the Cop$^+$-FND by the Kupffer cells may reduce the bioavailability of the FND and could explain the low amount of Cop$^+$-FND detected in the tumor after i.v. injection in the two mice used for the biodistribution study. Considering these results, to test the in vivo efficacy of Cop$^+$-FND:siAS (siRNA targeting *EWS-FLI1* oncogene) in animals we opted for intratumoral administration. Indeed, compared to i.v. administration, it was shown for other nanomedicines that i.t. injection renders a much larger tumor/tissue ratio in nanoparticles content,[32] thanks to their very low leakage to other organs. This property, combined to the fact that the siRNA sequence we use is directed towards the junction of *EWS-FLI1* oncogene that are only present in the tumor, make off-target side effects very unlikely.

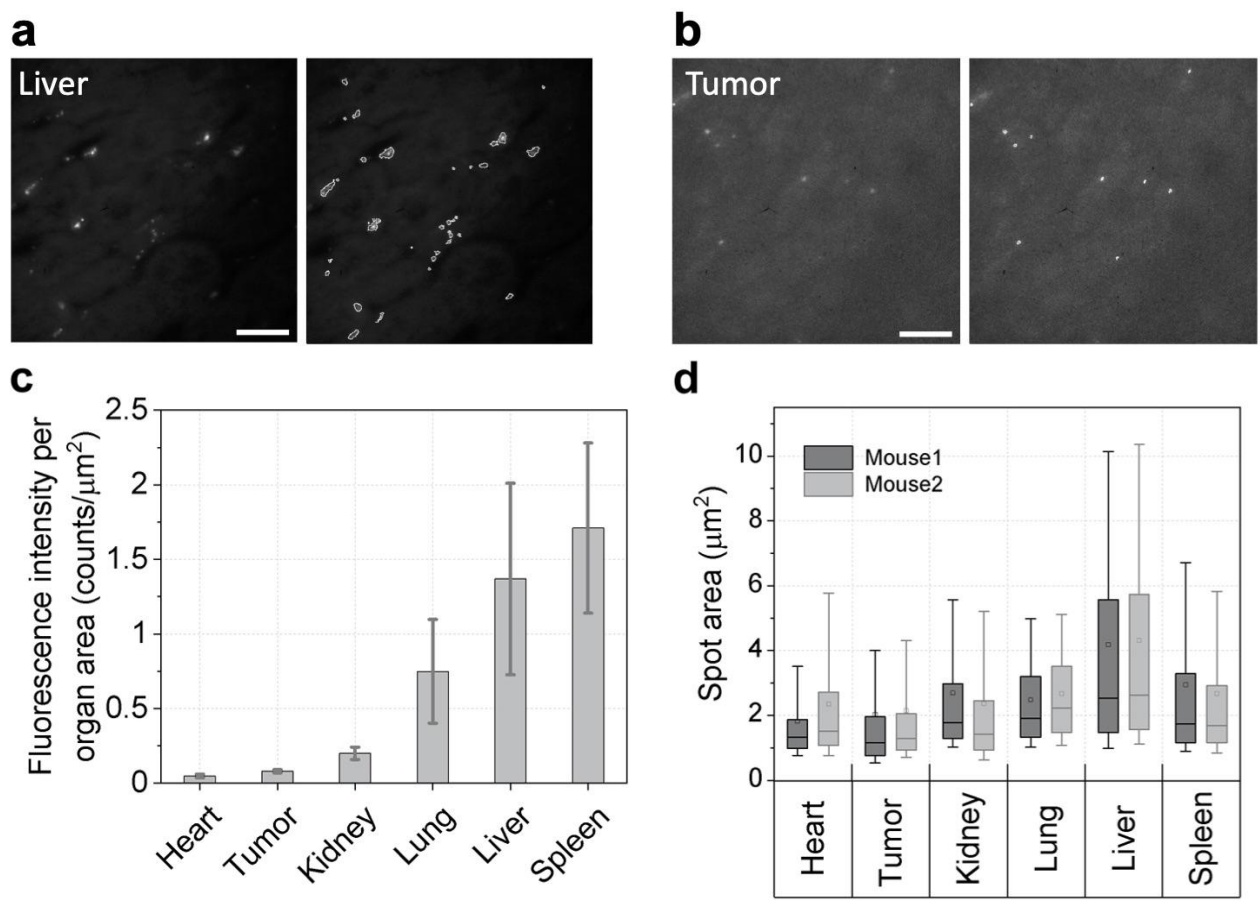

**Figure 3**. **Automatic quantification of the amount of Cop$^+$-FND and aggregation state in the different organ sections inferred from time-gated fluorescence imaging.** a,b) One fluorescence FOV of a liver (a) and tumor (b) section without (left) and with (right) the ROI surrounding automatically detected FND. Scale bar: 20 µm. c) Total fluorescence intensity of all ROI detected divided by the surface of the organs. d) Distribution of the ROI areas in the different organs. The boxplot characterizes a sample using the 25$^{th}$, 50$^{th}$ and 75$^{th}$ percentiles, and 90% and 10% whiskers; horizontal line in the box: median of the distribution; square: mean of the distribution. The outliers were not presented here. Data from *n*=2 organ sections from 2 mice.

### 2.2.4. FND cellular localization in liver histological sections.

To investigate in which compartments of the liver Cop$^+$-FND end up, we then took advantage of high magnification of individual image constituting the mosaic and of the bright field acquisition of HES stained sections. Fig. 2e shows examples of Cop$^+$-FND ROI superimposed to HES images. FND were found in majority internalized in endothelial cells bordering the sinusoids, and second, most probably inside Kupffer cells. The latter are liver-resident macrophages part of the mononuclear phagocyte system and essential to maintain liver homeostasis and scavenge xenobiotics and cell debris[29]. A staining specific of macrophages like F4/80[30] would allow to unambiguously distinguish Kupffer cells from endothelial cells, so that we could have an accurate quantification of each cell type. The capture by Kupffer cells, if it is confirmed, would not be favourable to the delivery of nanomedicine into the target tumour, but it is not the sole mechanism responsible for the low yield of nanoparticles ending up in tumor after systemic administration, as recently reported in models where Kupffer cells were removed[31]. The interaction with other organs than the liver, like the spleen, need also to be tuned to achieve a larger tumor delivery.

Let us further point out that the phagocytosis of nanodiamonds by Kupffer cells was already reported a decade ago[10], and that these cells are expected to play an important role in the elimination of these particles from the organism. However, we did not detect any nanodiamond in hepatocytes 24 h after administration, which indicates that we either did not wait long enough for their passing from the Kupffer cells to the hepatocyte before reaching the bile duct and being eliminated in the feces, or that they are sequestered for a longer period and might not be eliminated at all. In any case, the important phagocytosis of the $Cop^+$-FND by the Kupffer cells may reduce the bioavailability of the FND and could explain the low amount of $Cop^+$-FND detected in the tumor after i.v. injection in the two mice used for the biodistribution study. Considering these results, to test the in vivo efficacy of $Cop^+$-FND:siAS (siRNA targeting *EWS-FLI1* oncogene) in animals we opted for intratumoral administration. Indeed, compared to i.v. administration, it was shown for other nanomedicines that i.t. injection renders a much larger tumor/tissue ratio in nanoparticles content,[32] thanks to their very low leakage to other organs. This property, combined to the fact that the siRNA sequence we use is directed towards the junction of *EWS-FLI1* oncogene that are only present in the tumor, make off-target side effects very unlikely.

## 2.3. *In vivo* EWS-FLI1 inhibition efficacy by $Cop^+$-FND:siAS injected intratumorously

For this *in vivo* efficacy experiment, we treated 6 athymic mice bearing A673 subcutaneous tumors with $Cop^+$-FND:siAS complexes or with $Cop^+$-FND:siCt bearing a control, non-specific, siRNA. The mass of siRNA injected was 30 µg in both cases. We used the same FND:siRNA preparation (mass ratio 25:1) as the one applied to *in vitro* experiment that showed high inhibition efficacy (Fig. 1c). In the *in vivo* experiment, mice were sacrificed 48 h post-FND intratumoral administration and the expression of *EWS-FLI1* oncogene in tumors was measured by RT-qPCR. We observed a 28-fold decrease of *EWS-FLI1* mRNA levels in tumors treated with $Cop^+$-FND:siAS with respect to the tumors treated with siCt (Fig. 4a, $p=1.7\times10^{-8}$), indicating that $Cop^+$-FND efficiently delivered and released active siRNA.

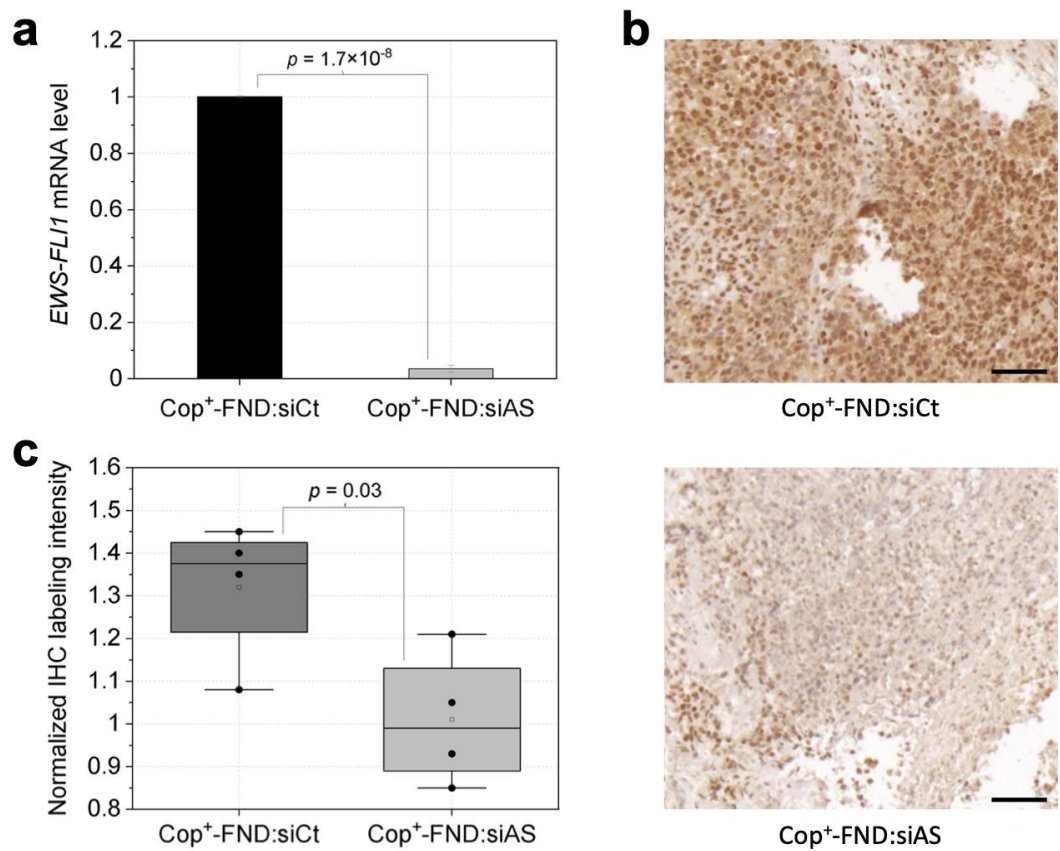

**Figure 4.** *In vivo* efficacy of Cop[+]-FND:siRNA in subcutaneous A673 xenografts 48 h after intratumoral administration. Cop[+]-FND:siRNA mass ratio of 25:1. a) *EWS-FLI1* mRNA expression levels in tumors treated either with a siRNA directed towards *EWS-FLI1* (siAS) or with the control siRNA (siCt); *n*=6 mice per group. b) Examples of bright field microscopy images of IHC preparations (siCT and siAS treatments) used for the quantification of EWS-FLI1 protein expression. Scale bars: 50 µm. c) Quantification of EWS-FLI1 IHC cell nucleus intensity from 4 mice (one tumor section per mouse) in each group, displayed as boxplots: the middle bar is the median, the empty square is the mean, and circles are values of each section.

Moreover, we investigated histologically the tumors and observed that Cop[+]-FND:siAS treated tumors had a lower ($p$=0.030) nuclear EWS-FLI1 immuno-histo-chemical (IHC) staining than Cop[+]-FND:siCt (Fig. 4b,c), except for one preparation over five which displayed an abnormal IHC labeling intensity distribution (Fig. S3). These results support the fact that Cop[+]-FND:siAS is able to efficiently deliver siRNA and inhibit a target gene *in vivo*. We also observed that 48 h after treatment, the proliferation marker Ki67 has a reduced expression (Fig. S5a,b) in the tumor of animals having received Cop[+]-FND:siAS and that their tumor volumes were significantly smaller ($p$=0.025) than in the control group (Fig. S5c). Analysis on longer durations would be necessary to assess a therapeutic efficacy. Furthermore, additional surface engineering of the particles can provide stealth effect to escape macrophage recognition. Among the most promising strategies is a controlled adsorption of functional proteins/peptides (hydrophobins, peptides designed from human CD47…) forming an "artificially designed" biomolecular corona[33].

We recently reported such potential for 7 nm sized detonation nanodiamonds (DNDs), rendered cationic by hydrogenation, and which can deliver after intravenous injection an efficient siAS to mice bearing the same xenografted Ewing sarcoma tumor[34]. We observed a two-fold decrease (50% inhibition) of *EWS-FLI1* mRNA levels in tumors of mice treated with DND:siAS compared to mice treated with DND:siCt. As one could have expected, this efficacy is much lower than the one we achieved here by intratumoral injection. In the same work using DND, we also investigated the nanodiamond tissue distribution after their radiolabeling with tritium (in replacement of hydrogen) and i.v. injection. A similar distribution as for Cop[+]-FND:siRNA was found, with the liver first, then the lungs and the spleen being the organs with the largest concentration of nanodiamonds. DND were also found in the feces but not in urine, which led us to the hypothesis that they could be eliminated in the bile through liver filtration. However, the tritiated DND could only provide whole organ-scale measurements and we could not spatially resolve their cellular fate, contrary to the experiments reported here using individually traceable FND (Fig. 2e).

# 3. Experimental

## 3.1. Cop$^+$-FND synthesis

Commercially available diamond nanocrystals (MSY 0-0.05, Microdiamant, Switzerland) obtained from milling High Pressure High Temperature nanodiamonds (type Ib, with nitrogen impurity concentration of ≈200 ppm) were oxidized by air in a furnace at 510 °C for 4 h and treated with a mixture of HF and HNO$_3$ (2:1) at 160°C for 2 days. The particles were washed with water, 1 M NaOH, 1 M HCl and water. NV centers were created in the nanodiamonds according to the published protocol[35]. Briefly, NDs were irradiated with a 16.6 MeV electron beam to obtain a high density of vacancies, annealed at 900°C (1 h) to create NV centers and oxidized by air (510°C for 4 h). The resulting powder was again treated with a mixture of HF and HNO$_3$ and then washed with NaOH, HCl and water. This procedure eventually provided an aqueous colloidal solution of fluorescent ND. FND colloid (60 mL, 2 mg/mL) dissolved in purified water (Milli-Q, Millipore, USA) was used for coating with methacrylate-terminated thin silica shell (hydrated SiO$_2$)[36,37] using a modified Stöber procedure[35]; amount of all components needed for silication was linearly recalculated to amount of FND (120 mg). Terminal methacrylate groups of the silica shell were used to grow a dense layer of statistical copolymer poly{(2-dimethylaminoethyl methacrylate)-*co*-[N-(2-hydroxypropyl)methacrylamide]} (poly(DMAEMA-*co*-HPMA)) via radical polymerization. HPMA and 2,2′-azobis(2-methylpropionitrile) (AIBN) were freshly recrystallized prior to use according to the previously described protocol[22,38]. Both monomers DMAEMA (1968 mg) and HPMA (656 mg) were dissolved in DMSO (7.5 mL); AIBN (752 mg) was added to the mixture that was then filtered using a 0.2 μm PTFE filter. Methacrylate-terminated FNDs in DMSO (120 mg, 752 μL) were added and the stirred mixture was secured in argon (three cycles vacuum – argon, 1 min – 1 min). The polymerization was performed for 3 days under argon at 55°C. The reaction was stopped by MeOH addition and two-step centrifugations were performed (20,000 g, during 20 min followed by 40,000 g during 20 min). The resulting Cop+-FND samples were centrifuged five times and always washed with nuclease-free water in an RNase-free flow box to ensure sample purity and sterility, providing the final concentration of 9 mg/mL. The samples were stored at -20 °C in MeOH for long-term storage or in nuclease free water at 4°C for biological and stability testing. Further characterization of the Cop$^+$-FND including $^1$H NMR spectra and thermogravimetric analysis are reported in Ref.[22].

### 3.2. siRNA sequences and Cop$^+$-FND:siRNA complexation

We ordered two custom MISSION® siRNA desalt from Sigma Aldrich (USA): siRNA directed toward *EWS-FLI1* (siAS), and a control irrelevant siRNA (siCt). The siAS sense strand sequence was: GCA GCA GAA CCC UUC UUA Ud(GA); the siAS antisense strand was : AUA AGA AGG GUU CUG CUG Cd(CC)). The siCt sense strand was: CGU UAC CAU CGA GGA UCC Ad(AA); the siCt antisense strand: UGG AUC CUC GAU GGU AAC Gd(CT)). All siRNA stock solutions were handled in RNase free water, pH 5.1. Before complexation with siRNA, the Cop$^+$-FND stock solution (9 mg/mL) was sonicated for 15 s in an ultrasonic bath. To obtain the selected Cop$^+$-FND:siRNA mass ratios (5:1, 10:1, 15:1, 20:1, 25:1, 30:1, 35:1, and 40:1), the corresponding amounts of Cop$^+$-FND stock solution were diluted by RNase free water resulting in a final volume of 22.2 μL and added into 2.8 μL (100 μM, 0.28 nmol) of siAS solution. The resulting solutions of Cop$^+$-FND:siRNA were gently sonicated 5-10 s in the ultrasonic bath, maintained at room temperature for 20 min and gently sonicated again. In a first approximation, the success of the complexation step was characterized by a clear and translucent suspension (Supporting Fig. S4), while a failure resulted in a flocculating milky dispersion. The concentration of Cop$^+$-FND:siRNA and the loading efficiency of siRNA was determined by spectrofluorimetry. To quantify the amount of free siRNA, the sample was centrifuged for 20 min at 20,000 *g* and the supernatant was analyzed using Qubit miRNA assay kit (ThermoFisher Scientific, USA) relative to control (particle-free). The total concentration of siRNA in the final solution prepared to be injected (25:1 mass ratio) was 1.5 mg/mL and 98 % of the siRNA was bound to Cop$^+$-FND.

### 3.3. Nanoconjugates colloidal stability as evaluated by DLS and ELS. TEM measurements.

*Z*-average diameter and apparent ζ-potential were measured with a Zetasizer Nano ZSP (Malvern Instruments, UK) equipped with a He-Ne laser emitting a power of 10 mW at 633 nm wavelength. Scattered light was collected at the backscatter angle of 173°. Mutual diffusion cefficient was obtained from a 3$^{rd}$-order cumulant analysis (Zetasizer Software 7.11, Malvern Instruments, UK) of $G^{(2)}(\tau)$-1, where $G^{(2)}(\tau) = <I(t)I(t+\tau)>_t$ ($<\cdot>_t$ denoting time average) is the second-order time intensity autocorrelation function and Z-average diameter was inferred from this fit using the Stokes-Einstein equation. Size distributions shown in Supporting Fig. S1c result from the non-negative least-squares analysis (General Purpose algorithm - regularizer 0.01) of Zetasizer Software 7.11. Unless stated otherwise, the measured Cop$^+$-FND samples were diluted by UltraPure™ DNase/RNase-Free Distilled Water (Nuclease free water, Thermo Fisher Scientific, pH 5.1) at 25°C with the final volume 0.6 mL (disposable cuvette); dynamic viscosity was set at 0.887 cP. Each sample was measured two times with an automatic duration; reported size diameters represent an average value of these measurements. Cop$^+$-FND:siAS sample tested in DMEM, 10% FCS (DMEM cell culture media) and 100% FCS at 37°C with the final volume 60 μL (quartz cuvette ZEN2112) was measured ten times (total duration ≈20 min); dynamic viscosity of the DMEM, 10% FCS (DMEM cell culture media) and 100% FCS at 37 °C was set at 0.726[39], 0.738[39] and 0.861 cP[40], respectively; dispersant refractive index was always set at 1.33 for all tested solvents.

ELS with a phase analysis of scattered light at forward angle of 13° was used for determination of the apparent ζ-potential. The samples were measured in nuclease-free water (pH 5.1) using a disposable cuvette (final volume 0.6 mL) with a dip cell and the following setting: monomodal analysis, two measurements with twenty sub-runs. To calculate the apparent ζ-potential value from Henry's equation[41] we have used Smoluchowsky approximation assuming value of the Henry's function 1.5[41,42]. For both types of measurement, the particle concentration was below 0.2 mg/mL; dilution of the solvent by testing sample was always lower than 5%.

TEM experiments were performed with a JEOL JEM-1011 electron microscope operated at 80 kV equipped with a Tengra (EMSIS, Germany) bottom-mounted camera. Particles were placed on copper grids (Structure Probe, Inc., USA) with a homemade Parlodion membrane and carbon coating. The grid was pretreated with a droplet of poly(ethylenimine) solution (MW=2.5 kDa, 0.1 mg/mL).[43] The droplet was removed with a piece of tissue after 10 min of incubation, the grid was placed on a droplet of deionized water for 1 min and dried once again.

### 3.4. In vitro inhibition of *EWS-FLI1* oncogene expression

Human Ewing carcinoma cells A673 were seeded at 2.5 to 5.10$^4$ cells/mL on a 24-wells plate or 6-wells plate (proliferation assay and RT-qPCR assay, respectively) in DMEM medium (Gibco) containing 10% bovine calf serum and 1% penicillin/streptomycin (Gibco) and incubated at 37°C, 5% $CO_2$ in moistly atmosphere one day before treatment. The medium was then replaced by 450 μL of same medium plus 50 µL of 10 mM HEPES pH 7.2, 100 mM NaCl containing Cop$^+$-FND bound to either siAS or siCt (100 nM per well). The Cop$^+$-FND:siRNA mass ratio was 25:1. A positive control was also tested, by using Lipofectamine 2000 (Invitrogen, USA) to deliver siRNA. Transfection was conducted in serum-reduced medium (Opti-MEM, Gibco, USA), during the first 4 h of incubation before to be replaced by DMEM containing 10 % bovine calf serum and penicillin streptomycin.

To determine the cellular proliferation, the cells were counted 72h post-treatment comparing the different treatments. For the oncogene inhibition measurement cells were incubated for 48 h to extract the total RNA by RNeasy RNA extraction kit (Qiagen, Spain), according to manufacturer's instructions. The total extracted RNA was dissolved in 10 µL of water and RNA concentration was determined by spectrophotometry at the wavelength of 260 nm (Nanodrop, Thermo Fisher Scientific, USA). The reverse transcription was performed with 3 µg RNA using High-capacity cDNA reverse transcription kit (ThermoFisher Scientific, CA, USA) according to the manufacturer's instructions. PCR quantification was carried out with 1 µL of cDNA using SYBR-green PCR Master mix (Applied Biosystems, ThermoFisher Scientific). In detail, the *EWS-FLI1* gene was amplified with the *EWS-* forward primer: 5'-AGC AGT TAC TCT CAG CAG AAC ACC -3' and *FLI1*-reverse primer: 5'-CCA GGA TCT GAT ACG GAT CTG GCT G-3' (Invitrogen). over 40 cycles, in a thermal cycler equipment (7500 Real-Time PCR system, Applied Biosystems, USA), as follows: 20 s at 95°C, followed by 40 cycles of 95°C for 15 s, 60°C for 30 s. The human *GAPDH* gene was used as control (forward primer: 5'- ACC CAC TCC TCC ACC TTT GAC -3', and reverse primer: 5'- CAT ACC AGG AAA TGA GCT TGA CAA-3'). Analysis of the expression levels of the genes of interest using QbasePLUS 2.0 software (Biogazelle, Belgium) and comparative CT (threshold cycle) methods was used to normalize the target CT by the *GAPDH* control gene CT (2e−ΔΔCT)[44].

### 3.5. Time-gated fluorescence microscope

The automated slide scanner setup relies on a first generation of the Pathfinder™ (Imstar S.A., France) slide scanner, composed of an epifluorescence inverted microscope (Eclipse Ti-E, Nikon, Japan) with a customized motorization (Imstar). For the fluorescence illumination in a pulse regime, we built our own laser system based on a modulated laser diode at 1064 nm wavelength amplified in a Nd$^{3+}$:YVO$_4$ crystal, following our previously published design[45]. We slightly modified this design to achieve sub-nanosecond (≈100 ps) pulse duration. Before being injected into the microscope, the output beam is first frequency doubled using a non-critical phase matching lithium triborate crystal (LBO-405, Eksma, Lithuania) placed in an oven (NCPM-405, Eksma) set at a temperature of 149°C, for the efficient nonlinear production of 532 nm wavelength light. Then, the laser beam is ×10 enlarged by a beam expander, reflected by a dichroic mirror (DM, z532rdc, Semrock, USA) and focused to the back focal plane of the microscope objective (oil immersion, Nikon Plan Apo ×60, numerical aperture 1.40), to produce a ≈100 µm diameter illumination spot. The fluorescence from FND is detected through the DM and after the elimination of residual excitation laser with a bandpass filter (697/75 nm, Semrock).

For the time-delayed imaging we use an intensified CCD array detector (ICCD PI-MAX3-1024i-18 mm-Gen-III-HBf-P46, Princeton Instrument, USA) which maximal 1 MHz internal clock frequency (SyncMaster output) serves as the master for the time-delayed recording. Imstar Pathfinder™ software was adapted to drive this ICCD and pass it the desired parameters. The latter include the delay between the clock front edge and the intensifier gate opening (set to 15 ns), the gate duration (110 ns), the intensifier gain (set to 5) and the number of gates before the detector reading (corresponding to a total pausing duration of 350 ms). Considering the delays in the cables and laser drivers electronic circuit, each ICCD clock front edge synchronization signal is delayed (and also reshaped in 50-ns TTL pulses) with a pulse Pulse Generator Delay apparatus (TGP110, Aim-TTi, UK, not displayed on Fig. 2a) so that it triggers a laser pulse exactly one period later (*i.e.*, 1 µs later).

### 3.6. Cop$^+$-FND:siRNA organ distribution

Nude mice were injected subcutaneously with 5.10$^6$ A673 cells, on one flank, and treatment was not begun until the tumor had reached a volume of 200 mm$^3$. After reaching this point, 200 µL of Cop$^+$-FND:siRNA, according to the protocol described beforehand, were injected intravenously in the mouse's caudal vein with a final siRNA concentration of 1 mg/kg and a Cop$^+$-FND:siRNA mass ratio of 25:1. Mice were sacrificed 24 h after the injection, by cervical dislocation in accordance with ethical procedures. Heart, lungs, liver, spleen, kidneys and tumor were collected and stored in FineFIX solution (Milestone srl, Italy), then embedded in paraffin and sectioned (thickness ≈4 µm). Sections were either let without any staining or colored with Hematoxylin/Eosin/Saffron (HES) solution. Sections of organs were then observed under a 60× magnification and 1.4 numerical aperture oil immersion microscope objective from the "home-made" time-gated fluorescence microscope.

Fluorescence and bright-field image of the entire organ were acquired under the form of a mosaic of fields of view, of size ≈110×110 µm each. The pausing duration for fluorescence was 350 ms, with a photocathode gain of 5, and a laser excitation power of 25 mW. The mosaic of the entire organ section was reconstructed using Imstar Pathfinder software. It was then possible, by zooming into the reconstructed image, to observe the whole organ section at subcellular resolution. Finally, the motorized microscope stage accommodates a 4-slides charger mount so that we could launch the fully automated acquisition of 4 organ sections at once. Overall, we could acquire an entire organ section (size between 25 and 110 mm$^2$) in 6 to 10 hours.

### 3.7. Automatic image processing for FND identification

To detect FNDs, we first applied a Tophat transformation to the fluorescence signal image; this procedure effectively removes the background inhomogeneities. In the second step we measured the variation of each pixel intensity value below the image median value was calculated from the Tophat image; the detection threshold was set as the median plus the standard deviation multiplied by a fixed factor. The ROIs generated by the previous step were filtered according the following criteria, and only the ROIs passing all the filters were retained:

(i) Global contrast: Value of the ratio between the ROI's average fluorescence intensity and the fluorescence intensity mode of the whole field of view. This parameter can be associated to the contrast of the ROI compared to the global image. The ROIs were considered as FNDs candidates only within a certain range of global contrast values;

(ii) Local contrast: ratio of the ROI's average fluorescence intensity to that of a small ring (few pixels) around the ROI. This allowed to eliminate the ROIs with sufficiently high average fluorescence level but low local contrast; these were most probably false positives;

(iii) Value of the total (integral) intensity of the object detected: below a specific threshold value, ROIs were considered as false positives;

(iv) Area of the ROI: set to prevent the detection of a large part of the organ. In lung for instance, some parts of the organ were highly fluorescent and could be detected as FNDs if no size filter was applied;

(v) Ratio of the fluorescence intensity between the first bleaching step and the acquisition step. When this ratio was within a specific range, the detected ROIs were considered as FNDs (the fluorescence of FND does not decrease after the bleaching step).

The values of these filter parameters are provided in the Supporting Table S1.

### 3.8. *In vivo* efficacy

Six-wee-old female nude mice (*n*=20) were subcutaneously injected with 5×10$^6$ A673 cells in the right dorsal back. Tumor volume is calculated as ($D$×$d^2$)/2, where $D$ and $d$ are the largest and smallest dimensions respectively, measured with a caliper as previously described[46]. Mice were treated when tumors reached approximately 75 mm$^3$. They were randomized at that time in treatment groups and administered a single injection. The Cop$^+$-FND:siRNA complexes (FND:siAS or FND:siCt) were assembled just before the intratumoral (i.t.) administration of 20 µL of Cop$^+$-FND:siRNA ready-to-be-injected solution (cf. §3.2), containing a mass of 30 µg of siRNA. Animals were euthanized 48 h post-administration and tumors were processed for either frozen for later qPCR assays or fixed in 4% paraformaldehyde and embedded in paraffin for histological analysis. Both siRNA directed toward *EWS-FLI1* (same sequences as for the *in vitro* inhibition experiment) and control siRNA directed toward *FLUC2* (Forward primer: 5'-GGA CGA GGA CGA GCA CUU CUU-3'; reverse primer: 5'-GAA GUG CUC GUC CUC GUC CUU-3') were produced by Sigma Aldrich.

### 3.9. Animal experiments ethical statement

All animal procedures were performed in accordance with the Guidelines for Care and Use of the Animal Facility of Institut Gustave Roussy (IGR, Villejuif, France) and of the Singular Scientific and Technical Infrastructures NanoBioSys$^‡$ U20 *in vivo* experimental platform of Vall d'Hebron Research Institute (VIHR, Barcelona, Spain). Biodistribution experiments were approved by IGR Animal Ethics Committee and the *in vivo* efficacy study was authorized by VHIR Animal Experimentation Ethical Committee (Ref.OH/9461/2).

### 3.10. Immunohistochemistry

The presence of EWS-FLI1 in tumor sections was analyzed by pre-treating paraffin-embedded formalin fixed sections with 100 mM citrate buffer (pH 6) in a pressure cooker. Sections were incubated with 10 % normal goat serum (NGS) in antibody diluent (1% BSA in Tris buffer 1X) and then of 1:50 dilution of anti-EWS-FLI1 antibody (MBS 300723, MyBiosource Inc., CA, USA). Secondary antibody consisted in a HRP conjugated system (EnVision+ System-HRP Labeled Polymer anti-rabbit from Agilent Dako, CA, USA), which was later visualized with diaminobenzidine colorimetric reagent solution (Agilent Dako) and counterstained with Harris hematoxylin (Sigma-Aldrich, MO, USA).

### 3.11. IHC quantification of EWS-FLI1

The quantification of EWS-FLI1 protein level in immuno-histochemistry tumor sections was done using a pipeline from the Pathfinder software (IMSTAR S.A., Paris). The first step consists in color calibration of RGB images: the experimentalist selects about 10 positive (P) and negative (N) nuclei; in each image the software calculates at each pixel the optical density (OD, decimal logarithm of the ratio of a nucleus pixel intensity to the intensity of the non-nucleus background) of each color ($OD_R$, $OD_G$ and $OD_B$, for red, green and blue respectively), normalizes it by $(OD_R^2+ OD_G^2+ OD_B^2)^{1/2}$, and then extract the medians for N and P nuclei, resulting in 6 values: $C^{P,N}_R$, $C^{P,N}_G$ and $C^{P,N}_B$ for P and N. This calibration is performed once for given batch of IHC. The second step consists in decomposing the RGB image in P and N monochrome components defined by as: $OD_R = C^P_R \times P + C^N_R \times N$, for R and P case, and similar expression for G and B. P and N values can be seen as OD associated to the ($C^P_R$, $C^P_G$, $C^P_B$) and ($C^N_R$, $C^N_G$, $C^N_B$) colors respectively. In the third step the nuclei are segmented in both P and N images using an automated thresholding followed by watershed-based separation and resulting in nuclei ROI. Finally, the mean value of P is calculated for all nuclei ROI and the average intensity over all nuclei ROI located within the region of the tumor outlined by the pathologist is used as the measure of EWS-FLI1 protein level.

### 3.12. Statistical comparisons

Unless otherwise stated, we used unpaired *t*-test to compare parameters between two conditions.

## Conclusions

We investigated the extension to *in vivo* of our pioneer *in vitro* experiments in which we showed that fluorescent nanodiamonds can transport siAS molecules inhibiting efficiently Ewing sarcoma *EWS-FLI1* oncogene in cultured cells[17]. We first evaluated the FND bioavailability after intravenous injection. To this aim, we took advantage of the intense emission of FND, their absence of photobleaching and their fluorescence lifetime which is one order of magnitude longer than the one of tissue autofluorescence, to build an automated imaging and detection pipeline. The latter provides high-resolution (≈120 Mpixels) fluorescence and histological images with single-particle and subcellular resolutions. Moreover, we also used an improved cationic functionalization of FND, that no more relies on physisorption of PEI[17,18], but on a cationic copolymer grafted covalently from FND surface forming Cop$^+$-FND nanoconjugates. This novel vector enabled to transport siAS efficiently into human Ewing Sarcoma cultured cells, and lead to a stronger inhibition of *EWS-FLI1* junction oncogene of 90% compared to the 55% previously obtained with physisorbed cationic polymer[17]. We then measured, with our custom imaging and quantification procedures, the main organ and tumor distribution of Cop$^+$-FND 24 h after i.v. injection of Cop$^+$-FND:siAS.

We found Cop$^+$-FND in all tissues investigated but with large differences of concentrations, and in particular a low one in the tumor. The dominant accumulation was in liver, spleen and lungs, as reported for other types of nanoparticles (administered intravenously to mouse for the same purpose) with similar size and surface charge[47], suggesting that this distribution might be driven by these basic physicochemical properties and not by chemical specificity. Our high spatial resolution imaging also allowed us to resolve individual Cop$^+$-FND from aggregated ones, and the lowest aggregation state was found in the tumor. This might reveal a preferential uptake by endocytosis, while we previously showed that micropinocytosis is the main pathway for the delivery of an active siRNA in the cytosol[18,23].

Based on these tissue bioavailability measurements, we decided for the *in vivo* treatment efficacy study to inject the Cop$^+$-FND:siAS directly in the tumor. We did observe, 48 h after injection, a large (28-fold) decrease of *EWS-FLI1* mRNA content in siAS treated tumor compared to control, translating in a significant decrease in EWS-FLI1 content in the tumor and an associated reduction of the tumor volume.

The method we developed and reported here to determine the tissue bioavailability of fluorescent nanodiamonds by automated time-delayed high-resolution fluorescence microscopy, can be extended to other luminescent nanoparticles with emission lifetimes up to ~100-200 ns, like near-infrared emitting quantum dots[48]. Longer lifetimes would require to increase the pulse laser repetition period beyond 1 μs, to avoid to filter-out legitimate photons. This would then require to increase the image pausing duration, in order to keep the photon counts constant, and the total acquisition time will increase beyond what is practically acceptable. Our method is therefore better suited for emission lifetime in a range of 50-200 ns. Finally, the ICCD could also be used in a dynamic mode to investigate the motion of FND-labelled subcellular compartments in live tissue sections, extending to thick samples studies carried-out in cultured cells[49].

## Author Contributions

Conceptualization, J.-R.B., A.P., P.C., I.A. and F.T.; methodology, S.C., M.K., A.P., Z.V.D.-R., J.-R.B., I.A., P.C. and F.T.; validation, S.C., M.K., Z.V.D.-R., J.-R.B., I.A., P.C. and F.T.; formal analysis, S.C., M.K., A.P., Z.V.D.-R., J.-R.B., and F.T.; investigation, S.C., M.K., A.P. and Z.V.D.-R.; resources, X.D., P.G., R.L.-A., O.T.M., M.K. and P.C.; software, A.P.; data curation, S.C. and J.-R.B.; writing—original draft preparation, S.C., J.-R.B. and F.T.; writing—review and editing, M.K., Z.V.D.-R., R.L.-A., O.T.M., J.-R.B., I.A., P.C. and F.T.; visualization, M.K., S.C., Z.V.D.-R. and F.T.; supervision, O.T.M., J.-R.B., I.A., P.C. and F.T.; project administration, O.T.M., J.-R.B., I.A., P.C. and F.T.; funding acquisition, X.D., O.T.M., J.-R.B., I.A., P.C. and F.T. All authors have read and agreed to the published version of the manuscript.

## Conflicts of interest

There are no conflicts to declare.

## Acknowledgements

We are grateful to Dr. Xavier Sanjuan from Bellvitge Hospital and to Dr. Paule Opolon from Gustave Roussy Institute for their help in anatomopathological analysis; to Olivia Bawa from Gustave Roussy Institute for sample preparation; to Sandra Mancilla and Natalia García-Aranda from Vall d'Hebron Institut de Recerca for their technical assistance in immunohistochemistry and *in vivo* qPCR assays, to Dr. David Chvatil from Institute of Nuclear Physics, Rez, for irradiation of nanodiamonds by electrons, and to Miroslava Guricova and Dr Helena Raabova for TEM measurements. This research was funded by the ERANET Euronanomed 2 program, through the French National Research Agency (ANR, grant number ANR-14-ENM2-0002 to F.T.) and through the Spanish Instituto de Salud Carlos III (ISCiii, grant numbers AC14/00032 to I.A. and AC14/00026 to O.T.M.); the Czech Science Foundation Project No.18-17071S (to P.C.); European Regional Development Fund, OP RDE, Projects: ChemBioDrug (No. CZ.02.1.01/0.0/0.0/16_019/0000729) (to P.C., M.K.) and CARAT (No. CZ.02.1.01/0.0/0.0/16_026/0008382) (to P.C.). Irradiations were supported through MSMT project No. LM2015056 within the CANAM infrastructure of the NPI CAS Rez.

## Notes and references

‡ [www.nanbiosis.es/portofolio/u20-in-vivo-experimental-platform](www.nanbiosis.es/portofolio/u20-in-vivo-experimental-platform)

1. S. Kunjachan, J. Ehling, G. Storm, F. Kiessling, and T. Lammers, *Chem. Rev.*, 2015, **115**, 10907–10937.
2. G. M. Thurber, K. S. Yang, T. Reiner, R. H. Kohler, P. Sorger, T. Mitchison, and R. Weissleder, *Nat. Commun.*, 2013, **4**, 1504.
3. A. Hoffman, H. Manner, J. W. Rey, and R. Kiesslich, *Nat. Rev. Gastroenterol. Hepatol.*, 2017, **14**, 421–434.
4. M. T. Matter, J. Li, I. Lese, C. Schreiner, L. Bernard, O. Scholder, J. Hubeli, K. Keevend, E. Tsolaki, E. Bertero, S. Bertazzo, R. Zboray, R. Olariu, M. A. Constantinescu, R. Figi, and I. K. Herrmann, *Adv. Sci.*, 2020, **7**, 2000912.
5. G. A. Roth, S. Tahiliani, N. M. Neu-Baker, and S. A. Brenner, *Wiley Interdiscip. Rev. Nanomedicine Nanobiotechnology*, 2015, **7**, 565–579.
6. E. D. SoRelle, O. Liba, J. L. Campbell, R. Dalal, C. L. Zavaleta, and A. de la Zerda, *Elife*, 2016, **5**.
7. K. Turcheniuk and V. N. Mochalin, *Nanotechnology*, 2017, **28**, 252001.
8. H.-C. Chang, W. W.-W. Hsiao, and M.-C. Su, *Fluorescent Nanodiamonds*, John Wiley & Sons, Ltd, Chichester, UK, 2018.
9. G. Gao, Q. Guo, and J. Zhi, *Small*, 2019, **15**, 1902238.
10. Y. Yuan, Y. Chen, J.-H. Liu, H. Wang, and Y. Liu, *Diam. Relat. Mater.*, 2009, **18**, 95–100.
11. M. Radtke, E. Bernardi, A. Slablab, R. Nelz, and E. Neu, *Nano Futur.*, 2019, **3**, 042004.
12. A. Hegyi and E. Yablonovitch, *Nano Lett.*, 2013, **13**, 1173–1178.
13. J.-J. Greffet, J.-P. Hugonin, M. Besbes, N. D. Lai, F. Treussart, and J.-F. Roch, *arXiv.org*, 2011, **quant-ph**, 1107.0502.
14. S. Coda, A. J. Thompson, G. T. Kennedy, K. L. Roche, L. Ayaru, D. S. Bansi, G. W. Stamp, A. V. Thillainayagam, P. M. W. French, and C. Dunsby, *Biomed. Opt. Express*, 2014, **5**, 515.
15. O. Faklaris, D. Garrot, V. Joshi, F. Druon, J. Boudou, T. Sauvage, P. Georges, P. A. Curmi, and F. Treussart, *Small*, 2008, **4**, 2236–2239.
16. T. Wu, Y.-K. Tzeng, W.-W. Chang, C.-A. Cheng, Y. Kuo, C. Chien, H.-C. Chang, and J. Yu, *Nat. Nanotechnol.*, 2013, **8**, 682–689.
17. A. Alhaddad, M.-P. Adam, J. Botsoa, G. Dantelle, S. Perruchas, T. Gacoin, C. Mansuy, S. Lavielle, C. Malvy, F. Treussart, and J.-R. Bertrand, *Small*, 2011, **7**, 3087–3095.
18. A. Alhaddad, C. Durieu, G. Dantelle, E. Le Cam, C. Malvy, F. Treussart, and J.-R. Bertrand, *PLoS One*, 2012, **7**, e52207.
19. M. Chen, X.-Q. Zhang, H. B. Man, R. Lam, E. K. Chow, and D. Ho, *J. Phys. Chem. Lett.*, 2010, **1**, 3167–3171.
20. S. Agarwal, Y. Zhang, S. Maji, and A. Greiner, *Mater. Today*, 2012, **15**, 388–393.
21. P. van de Wetering, N. M. E. Schuurmans-Nieuwenbroek, M. J. van Steenbergen, D. J. A. Crommelin, and W. E. Hennink, *J. Control. Release*, 2000, **64**, 193–203.
22. M. Kindermann, J. Neburkova, E. Neuhoferova, J. Majer, M. Guricova, V. Benson, and P. Cigler, *hal.archives-ouvertes.fr*, 2021.
23. J.-R. R. Bertrand, C. Pioche-Durieu, J. Ayala, T. Petit, H. A. Girard, C. P. Malvy, E. Le Cam, F. Treussart, and J.-C. C. Arnault, *Biomaterials*, 2015, **45**, 93–98.
24. O. Delattre, J. Zucman, T. Melot, X. Sastre Garau, J.-M. Zucker, G. M. Lenoir, P. F. Ambros, D. Sheer, C. Turc-Carel, T. J. Triche, A. Aurias, and G. Thomas, *N. Engl. J. Med.*, 1994, **331**, 294–299.
25. S. R. Hemelaar, A. Nagl, F. Bigot, M. M. Rodríguez-García, M. P. de Vries, M. Chipaux, and R. Schirhagl, *Microchim. Acta*, 2017, **184**, 1001–1009.
26. H. Maeda, H. Nakamura, and J. Fang, *Adv. Drug Deliv. Rev.*, 2013, **65**, 71–79.
27. S. Wilhelm, A. J. Tavares, Q. Dai, S. Ohta, J. Audet, H. F. Dvorak, and W. C. W. Chan, *Nat. Rev. Mater.*, 2016, **1**, 16014.
28. H. Maeda, *Adv. Drug Deliv. Rev.*, 2015, **91**, 3–6.
29. L. J. Dixon, M. Barnes, H. Tang, M. T. Pritchard, and L. E. Nagy, *Compr. Physiol.*, 2013, **3**, 785–97.
30. B. G. Lopez, M. S. Tsai, J. L. Baratta, K. J. Longmuir, and R. T. Robertson, *Comp. Hepatol.*, 2011, **10**, 2.
31. A. J. Tavares, W. Poon, Y. N. Zhang, Q. Dai, R. Besla, D. Ding, B. Ouyang, A. Li, J. Chen, G. Zheng, C.

Robbins, W. C. W. Chan, and C. J. Murphy, *Proc. Natl. Acad. Sci. U. S. A.*, 2017, **114**, E10871–E10880.
32. T. Lammers, P. Peschke, R. Kühnlein, V. Subr, K. Ulbrich, P. Huber, W. Hennink, and G. Storm, *Neoplasia*, 2006, **8**, 788–795.
33. V. H. Nguyen and B.-J. Lee, *Int. J. Nanomedicine*, 2017, **12**, 3137–3151.
34. S. Claveau, É. Nehlig, S. Garcia-Argote, S. Feuillastre, G. Pieters, H. A. Girard, J.-C. Arnault, F. Treussart, and J.-R. Bertrand, *Nanomaterials*, 2020, **10**, 553.
35. T. Rendler, J. Neburkova, O. Zemek, J. Kotek, A. Zappe, Z. Chu, P. Cigler, and J. Wrachtrup, *Nat. Commun.*, 2017, **8**, 14701.
36. J. Slegerova, M. Hajek, I. Rehor, F. Sedlak, J. Stursa, M. Hruby, and P. Cigler, *Nanoscale*, 2015, **7**, 415–420.
37. I. Rehor, H. Mackova, S. K. Filippov, J. Kucka, V. Proks, J. Slegerova, S. Turner, G. Van Tendeloo, M. Ledvina, M. Hruby, P. Cigler, G. Van Tendeloo, M. Ledvina, M. Hruby, P. Cigler, and G. Van Tendeloo, *Chempluschem*, 2014, **79**, 21–24.
38. J. Neburkova, M. Hajek, I. Rehor, J. Schimer, F. Sedlak, J. Stursa, M. Hruby, and P. Cigler, in *Methods in Pharmacology and Toxicology*, Humana Press Inc., 2017, pp. 169–189.
39. R. Ahmad Khanbeigi, T. F. Abelha, A. Woods, O. Rastoin, R. D. Harvey, M.-C. Jones, B. Forbes, M. A. Green, H. Collins, and L. A. Dailey, *Biomacromolecules*, 2015, **16**, 733–742.
40. O. Cakmak, C. Elbuken, E. Ermek, A. Mostafazadeh, I. Baris, B. Erdem Alaca, I. H. Kavakli, and H. Urey, *Methods*, 2013, **63**, 225–232.
41. G. V. Lowry, R. J. Hill, S. Harper, A. F. Rawle, C. O. Hendren, F. Klaessig, U. Nobbmann, P. Sayre, and J. Rumble, *Environ. Sci. Nano*, 2016, **3**, 953–965.
42. A. V. Delgado, F. González-Caballero, R. J. Hunter, L. K. Koopal, and J. Lyklema, *J. Colloid Interface Sci.*, 2007, **309**, 194–224.
43. I. Rehor and P. Cigler, *Diam. Relat. Mater.*, 2014, **46**, 21–24.
44. J. Hellemans, G. Mortier, A. De Paepe, F. Speleman, and J. Vandesompele, *Genome Biol.*, 2008, **8**, R19.
45. X. Délen, F. Balembois, and P. Georges, *Opt. Lett.*, 2014, **39**, 997.
46. M. Pesarrodona, T. Jauset, Z. V. Díaz-Riascos, A. Sánchez-Chardi, M. Beaulieu, J. Seras-Franzoso, L. Sánchez-García, R. Baltà-Foix, S. Mancilla, Y. Fernández, U. Rinas, S. Schwartz, L. Soucek, A. Villaverde, I. Abasolo, and E. Vázquez, *Adv. Sci.*, 2019, **6**, 1900849.
47. H. Elhamess, J.-R. Bertrand, J. Maccario, A. Maksimenko, and C. Malvy, *Oligonucleotides*, 2009, **19**, 255–264.
48. S. Bouccara, A. Fragola, E. Giovanelli, G. Sitbon, N. Lequeux, T. Pons, and V. Loriette, *J. Biomed. Opt.*, 2014, **19**, 051208.
49. S. Haziza, N. Mohan, Y. Loe-Mie, A.-M. Lepagnol-Bestel, S. Massou, M.-P. Adam, X. L. Le, J. Viard, C. Plancon, R. Daudin, P. Koebel, E. Dorard, C. Rose, F.-J. Hsieh, C.-C. Wu, B. Potier, Y. Herault, C. Sala, A. Corvin, B. Allinquant, H.-C. Chang, F. Treussart, and M. Simonneau, *Nat. Nanotechnol.*, 2017, **12**, 322–328.

# Supplementary Information

**Table S1.** Characterization of the colloidal properties of batch #1 of Cop$^+$-FND and Cop$^+$-FND:siAS complexes with a mass ratio 25:1 using DLS and ELS. Two first rows, all columns: mean ± standard deviation of two measurements. Rows 3 to 5 (complex in specific medium): the size was measured ten times (during approximately 20 min). As indicators of colloidal stability, two values are displayed: 1$^{st}$ / 10$^{th}$ measurements. More detailed information about the measurement conditions can be found in the Experimental section.

|  | Z-average diameter [nm] | Apparent ζ-potential [mV] | Electrophoretic mobility [μm·cm/V·s] | Conductivity [μS/cm] | Applied voltage [V] |
|---|---|---|---|---|---|
| **Cop$^+$-FND** (nuclease free water, 25 °C) | 138.0 ± 1.0 | 45.8 ± 2.9 | 3.59 ± 0.23 | 5.96 ± 0.05 | 9.5 |
| **Cop$^+$-FND:siRNA (siAS)** (nuclease free water, 25 °C) | 150.9 ± 6.2 | 38.4 ± 0.8 | 3.01 ± 0.07 | 18.30 ± 0.00 | 5.0 |
| **Cop$^+$-FND:siRNA (siAS)** (DMEM, 37 °C) | 157.9 / aggregated | - | - | - | - |
| **Cop$^+$-FND:siRNA (siAS)** (10% FCS, DMEM, 37 °C) | 150.4 / 152.9 | - | - | - | - |
| **Cop$^+$-FND:siRNA (siAS)** (100% FCS, 37 °C) | 229.9 / 287.7 | - | - | - | - |

**Table S2.** Ranges of the automatic detection parameters used to identify Cop$^+$-FND in function of the organ considered. These parameters need to be adjusted to each type of organ due to large differences in their autofluorescence intensities.

|  | Global contrast | Local contrast | Total intensity [counts/350 ms per ROI] | Area [μm$^2$] | | Ratio of intensity between bleaching and acquisition steps | |
|---|---|---|---|---|---|---|---|
|  | min | min | min | min | max | min | max |
| Lung | 1.3 | 1.2 | 30 | 0.25 | 7.5 | 0.65 | 1.25 |
| Heart | 0.8 | 1.27 | 20 | 0.4 | 9 | 0.75 | 1.25 |
| Tumor | 1.3 | 1.13 | 10 | 0.25 | 60 | 0.65 | 1.25 |
| Liver | 1.3 | 1.2 | 30 | 0.25 | 20 | 0.65 | 1.25 |
| Spleen | 1.3 | 1.2 | 30 | 0.25 | 20 | 0.65 | 1.25 |
| Kidney | 1.3 | 1.2 | 30 | 0.25 | 20 | 0.65 | 1.25 |

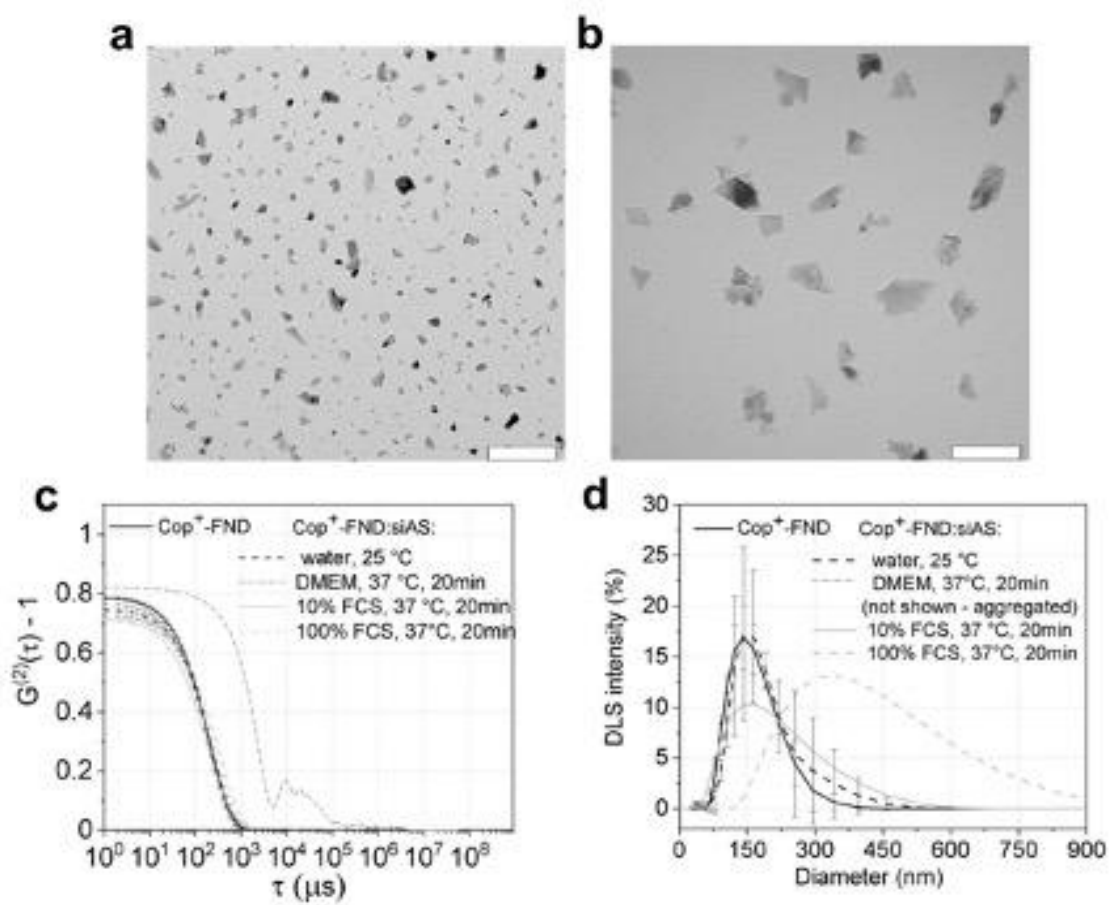

**Figure S1.** a, b) TEM images of FND samples. a) FND; scale bar: 200 nm. b) Cop[+]-FND, scale bar:100 nm. The samples were prepared on carbon-coated copper grids. c, d) Size distribution, as measured by DLS, of Cop[+]-FND and Cop[+]-FND:siAS of Table S1 samples (mass ratio 25:1). c) Raw DLS time intensity correlation functions $G^{(2)}(\tau)-1$. d) Intensity size distribution as inferred from the correlation data a), using non-negative least square analysis. In water at 25°C (dark grey dashed line) Cop[+]-FND:siAS size measurement was done twice and we report the average value ± standard deviation. Cop[+]-FND:siAS in DMEM alone aggregated after 20 min. For Cop[+]-FND:siAS in DMEM with 10% FCS (conventional cell culture media) and in 100% FCS we display only the 10[th] measurement (after 20 min).

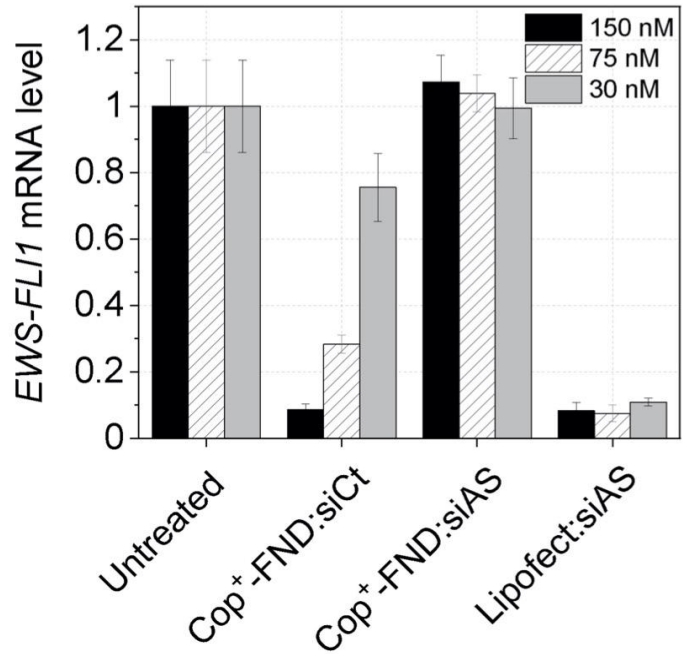

**Figure S2.** *EWS-FLI1* **oncogene inhibition** (measured by RT-qPCR) by Cop[+]-FND complexed with different concentration (30, 75, 150 nM) of siRNA AntiSense (siAS) directed against *EWS-FLI1* junction oncogene or Control (irrelevant siRNA, siCt), and compared to siAS complexed with Lipofectamine 2000 (Lipofect 2000, as a positive control, used without serum added to the medium), at different siRNA concentrations. Cop[+]-FND:siRNA mass ratio was kept constant and equal to 65:1. mRNA *EWS-FLI1* was extracted 24 h after incubation.

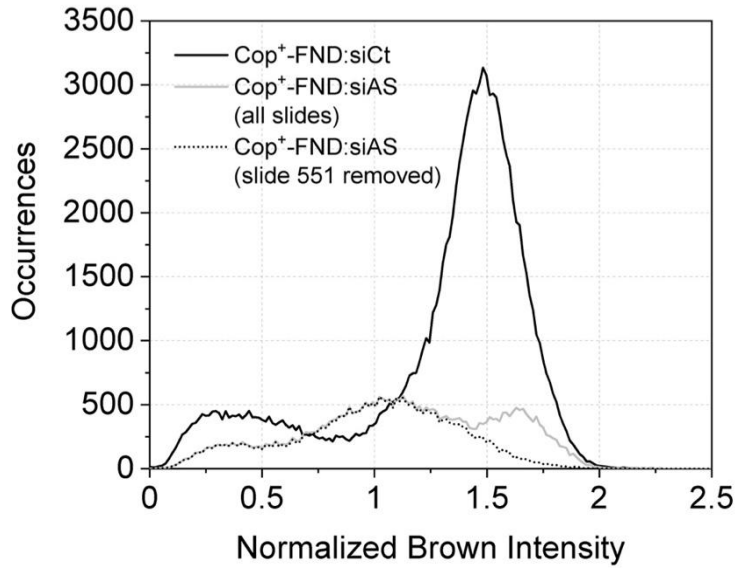

**Figure S3. Distributions of EWS-FLI1 immunostaining intensity within individual cell nucleus of tumor region selected by the histopathologist.** Two groups are displayed: animals treated by Cop[+]-FND:siCt (black solid line, data of 4 mice pooled) and Cop[+]-FND:siAS (grey solid line, data from 5 mice). In the case of Cop[+]-FND:siAS, the slide histological preparation #551 (1 mice) yielded a high staining intensity, larger than the one of Cop[+]-FND:Ct, which led us to suspect a labeling issue specific to this sample. After removing this problematic slide (black dashed line, 4 mice) we obtain similar shapes for Cop[+]-FND:siCt and Cop[+]-FND:siAS with two peaks, one at low intensity (around 0.3) corresponding to EWS-FLI1 negative labeling and the other one at high intensity (≈1.5 in the case of Cop[+]-FND:siCt) associated to positive nuclei. Cop[+]-FND to siRNA mass ratio of 25:1.

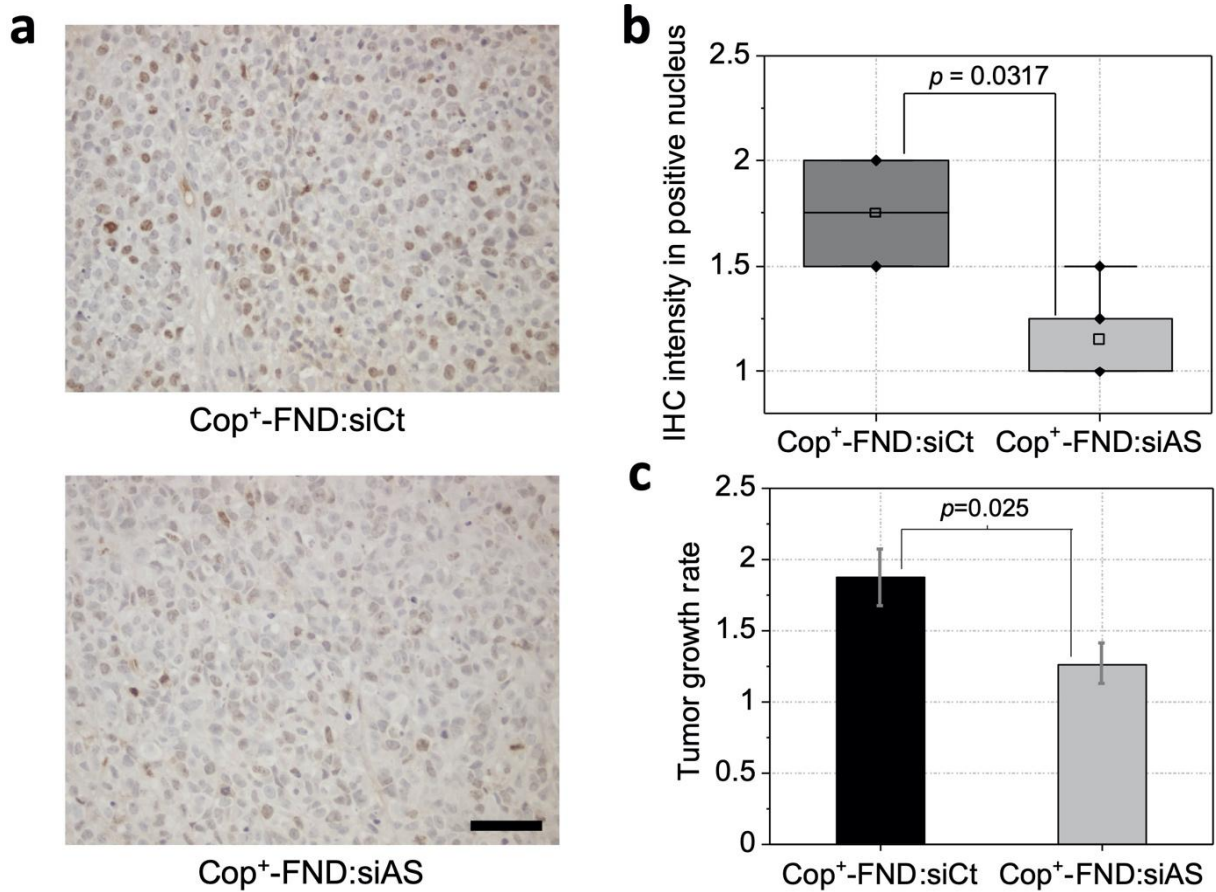

**Figure S4. *In vivo* efficacy of Cop+-FND:siRNA in subcutaneous A673 xenografts 48 h after intratumoral administration.** Cop+-FND:siRNA mass ratio of 25:1. a) Ki67 IHC in tumors treated with FND with control siRNA (Cop+-FND:siCt) or the siRNA targeting *EWS-FLI1* gene (Cop+-FND:siAS); *n*=4-5 mice per group; scale bar: 50 µm. b) IHC quantification corresponding to the scoring of Ki67 labeling intensity within cell nuclei by two blinded observers and showing statistically significant (*p*=0.0317, using Wilcoxon-Mann-Whitney test) lower Ki67 content in Cop+-FND:siAS treated sample compared to Cop+-FND:siCt. c) Tumor volume growth rate expressed as the ratio of the final volume to the initial one, 48 h after i.t. administration of Cop+-FND:siRNA. Treatement with siAS yielded a statistically significant (*p*=0.0255) smaller growth rate than with siCt.

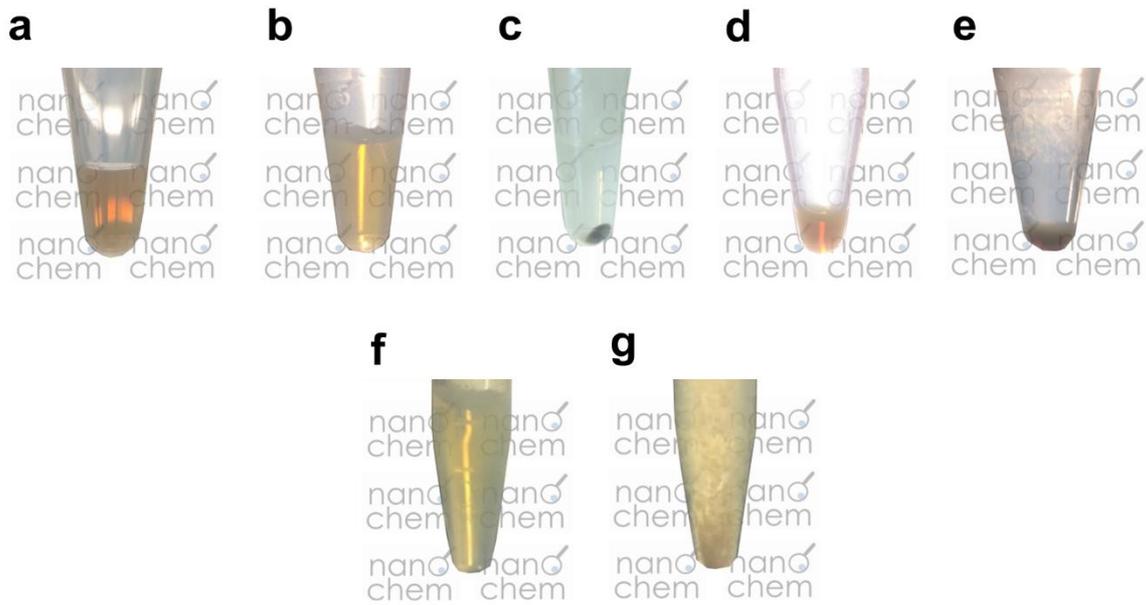

**Figure S5. Picture of dispersions** of a) Cop$^+$-FND (8 mg/mL); b) Cop$^+$-FND:siAS (3.9 mg Cop$^+$-FND/mL; 0.16 mg siAS/mL; i.e. 750 µg Cop$^+$-FND/tube; 30 µg siAS/tube); c) Sample (b) after centrifugation (20,000/ min); d) Dissolved sample (c) with the following concentration of components: 12.5 mg Cop$^+$-FND/mL; 0.5 mg siAS/mL, corresponding to the 25:1 mass ratio; e) Dissolved sample (c) with the following concentration of components: 37.5 mg Cop$^+$-FND/mL; 1.5 mg siAS/mL; f) FND (2 mg/mL), g) FND-PEI (0.8 kDa, branched) mixture (1:1 v/v; 2 mg FND/mL; 0.9 mg PEI/mL). Samples (a), (b), (d), (f) show the colloidal opalescence characteristic for sample stability. The opalescence of the sample (e) is masked due to high concentration of the complexes. Sample (g) exhibits aggregation which is typical for the mixture of FND and PEI but that can be reversed, resulting back in a transparent colloidal ND-PEI suspension (sample not shown).

\*\*\*